\documentclass[11pt]{article}
\usepackage{graphicx}

\begin{document}

\def\xslash#1{{\rlap{$#1$}/}}
\def \p {\partial}
\def \dd {\psi_{u\bar dg}}
\def \ddp {\psi_{u\bar dgg}}
\def \pq {\psi_{u\bar d\bar uu}}
\def \jpsi {J/\psi}
\def \psip {\psi^\prime}
\def \to {\rightarrow}
\def\bfsig{\mbox{\boldmath$\sigma$}}
\def\DT{\mbox{\boldmath$\Delta_T $}}
\def\xit{\mbox{\boldmath$\xi_\perp $}}
\def \jpsi {J/\psi}
\def\bfej{\mbox{\boldmath$\varepsilon$}}
\def \t {\tilde}
\def\epn {\varepsilon}
\def \up {\uparrow}
\def \dn {\downarrow}
\def \da {\dagger}
\def \pn3 {\phi_{u\bar d g}}

\def \p4n {\phi_{u\bar d gg}}

\def \bx {\bar x}
\def \by {\bar y}

\begin{center}
{\Large\bf  Transverse Momentum Dependent Factorization for Quarkonium Production at Low Transverse Momentum }
\par\vskip20pt
J.P. Ma$^{1,2}$, J.X. Wang$^{3}$ and  S. Zhao$^{1}$     \\
{\small {\it
$^1$ Institute of Theoretical Physics, Academia Sinica,
P.O. Box 2735,
Beijing 100190, China\\
$^2$ Center for High-Energy Physics, Peking University, Beijing 100871, China  \\
$^3$ Institute of High Energy Physics, Academia Sinica, P.O. Box 918(4), Beijing 100049, China
}} \\
\end{center}
\vskip 1cm
\begin{abstract}
Quarkonium production in hadron collisions at low transverse
momentum $q_\perp \ll M$ with $M$ as the quarkonium mass can be used
for probing transverse momentum dependent (TMD) gluon distributions.
For this purpose, one needs to establish the TMD factorization for
the process. We examine the factorization at the one-loop level for
the production of $\eta_c$ or $\eta_b$. The perturbative coefficient
in the factorization is determined at one-loop accuracy. Comparing
the factorization derived at tree level and that beyond the tree
level, a soft factor is, in general, needed to completely cancel
soft divergences. We have also discussed possible complications of
TMD factorization of p-wave quarkonium production. \vskip 5mm
\noindent
\end{abstract}
\vskip 1cm
\par

\par\vskip20pt
Quarkonium production in hadron collision can be used to explore the
gluon content of hadrons, because the quarkonium is dominantly
produced through gluon-gluon fusions. For the produced quarkonium
with large transverse momentum, one can apply QCD collinear
factorizations for long distance effects of the initial hadrons. In
this case, one can extract the standard gluon distributions(see,
e.g., \cite{MNS}). If the quarkonium is produced with small
transverse momentum $q_\perp$, it can be thought that the small
$q_\perp$ is generated at least partly from the transverse motion of
gluons inside the initial hadrons. In this case, one can apply
transverse momentum dependent(TMD) factorization for initial
hadrons. Therefore, the production with small $q_\perp$ allows
access to TMD gluon distributions.
\par
Factorizations with TMD quark distributions and fragmentation
functions have been studied intensively beyond tree level in
different processes in \cite{CS,CSS,JMY1,CAM}. In comparison, the
factorization with TMD gluon distributions beyond tree level has
only been studied for Higgs production in hadron collision in
\cite{JMYG}. Recently, the TMD factorization of quarkonium
production has been derived at tree level in \cite{BoPi}, and based
on it numerical predictions have been obtained. For theoretical
consistency and precision, it is important to examine the TMD
factorization beyond tree level. From early studies in
\cite{CS,CSS,JMY1,JMYG}, it is known that a soft factor needs to be
implemented into the factorization.  In this work, we examine TMD
factorization of $\eta_c$ or $\eta_b$ production at one-loop level.

\par

A quarkonium is dominantly a bound state of a heavy quark $Q$ and
its antiquark $\bar Q$. Because of the heavy mass the $Q\bar Q$ pair
is of a nonrelativistic system. To separate the nonperturbative
effects related to the quarkonium in its production, one can employ
nonrelativistic QCD (NRQCD) factorization \cite{nrqcd} by an
expansion of the small velocity  of $Q$ relative to $\bar Q$ . The
inclusive production of a quarkonium at moderate or large $q_\perp$
has been studied intensively both in theory and in experiments. In
the last five years, important progresses were made in the study of
the next-to-leading order QCD correction  for $J/\psi$ production in
hadron collisions \cite{NJ1} and power corrections \cite{KQS}. The
activities in this field can be seen in \cite{HQPPO}. It should be
noted that  in experiment it is also possible to study the inclusive
production at low $q_\perp$. For example, a $J/\psi$ produced at
LHCb can be measured with $q_\perp$ smaller than $1$GeV \cite{LHCb}.
Therefore, with theoretically established TMD factorization, one can
extract from experimental results TMD gluon distributions.
\par
\par
We will use the  light-cone coordinate system, in which a vector
$a^\mu$ is expressed as $a^\mu = (a^+, a^-, \vec a_\perp) =
((a^0+a^3)/\sqrt{2}, (a^0-a^3)/\sqrt{2}, a^1, a^2)$ and $a_\perp^2
=(a^1)^2+(a^2)^2$. We introduce two light cone vectors $ n^\mu =
(0,1,0,0)$ and $l^\mu =(1,0,0,0)$ and the transverse metric
$g_\perp^{\mu\nu}=g^{\mu\nu}-n^\mu l^\nu -n^\nu l^\mu$. We consider
the process
\begin{equation}
   h_A (P_A) + h_B (P_B) \to \eta_Q (q) + X,
\label{proc}
\end{equation}
in the kinematical region $Q^2 =q^2 \gg q^2_\perp$ with $Q=M_{\eta_Q}$ as the mass
of $\eta_Q$,
where $\eta_Q$ stands for $\eta_c$ or $\eta_b$. The momenta of the initial hadrons and of the quarkonium are given by
\begin{equation}
  P_A^\mu \approx (P_A^+,0,0,0), \ \ \  P_B^\mu \approx (0,P_B^-, 0,0), \ \ \  q^\mu =(x P_A^+, y P_B^-, \vec q_\perp),
\end{equation}
where we have neglected masses of hadrons, i.e., $P_A^- \approx 0$ and $P_B^+ \approx 0$. In the kinematic region
of $q_\perp \ll Q$ TMD factorization can be applied with corrections suppressed by positive powers of $q_\perp/Q$.
It is clear that in the kinematical region with $q_\perp \sim Q$ or $q_\perp \gg Q$ the TMD factorization can not be used.
In these regions one can use collinear factorization as studied in \cite{NJ1}.

\par
For each hadron in the initial state, one can define its TMD gluon
distribution. We introduce the gauge link along the direction $u^\mu
= (u^+, u^-,0,0)$,
\begin{equation}
{\mathcal L}_u (z,-\infty) = P \exp \left ( -i g_s \int^0_{-\infty}  d\lambda
     u\cdot G (\lambda u + z) \right ) ,
\end{equation}
where the gluon field is in the adjoint representation. At leading
twist, one can define two TMD gluon distributions through the gluon
density matrix \cite{TMDGP},
\begin{eqnarray}
&& \frac{1}{x P^+} \int \frac{ d\xi^- d^2 \xi_\perp}{(2\pi)^3}
e^{ - i x \xi^- P^+_A + i \vec \xi_\perp \cdot \vec k_\perp}
 \langle h_A \vert \left ( G^{+\mu} (\xi ) {\mathcal L}_u (\xi,-\infty) \right )^a
            \left ( {\mathcal L}_u^\dagger (0,-\infty) G^{+\nu}(0) \right )^a \vert h_A \rangle
\nonumber\\
  &&   =-\frac{1}{2} g_\perp^{\mu\nu} f_{g/A} (x,k_\perp, \zeta^2_u,\mu)
   + \left (k_\perp^\mu k_\perp^\nu + \frac{1}{2}g_\perp^{\mu\nu} k_\perp^2 \right ) h_{g/A} (x,k_\perp,
   \zeta^2_u,\mu),
\label{DEF}
\end{eqnarray}
with $\xi^\mu =(0,\xi^-,\vec \xi_\perp)$. $x$ is the momentum
fraction carried by the gluon inside $h_A$. The gluon has also a
nonzero transverse momentum $\vec k_\perp$. The definition is given
in nonsingular gauges. It is gauge invariant. In singular gauges,
one needs to add gauge links along transverse direction at
$\xi^-=-\infty$ \cite{TMDJi}.  Due to the gauge links, the TMD gluon
distributions also depend on the vector $u$ through the variable
\begin{equation}
\zeta^2_u = \frac {(2u\cdot P_A)^2}{u^2} \approx   \frac {2  u^-}{u^+} \left (P_A^+\right )^2.
\label{Zeta}
\end{equation}
In the definition, the limit $u^+ \ll u^-$ is taken in the sense
that one neglects all contributions suppressed by negative powers of
$\zeta_u^2$.
\par
From the definition in Eq. (\ref{DEF}), there are two TMD gluon
distributions. The distribution $f_{g/A}$ corresponds to the
standard gluon distribution in collinear factorization. The
distribution $h_{g/A}$ describes gluons with linear polarization
inside $h_A$. The relevant phenomenology of $h_{g/A}$ has been only
recently studied \cite{SecG1,SecG2,SecG3}. Through the process
studied here, one can also obtain information about this
distribution \cite{BoPi}. For $h_B$, one can also define two TMD
gluon distributions $f_{g/B}$ and $h_{g/B}$ similar to those in Eq.
(\ref{DEF}), in which the gauge links are along the direction
$v^\mu=(v^+,v^-,0,0)$ instead of $u$ and the limit $v^+\gg v^-$ is
taken. Therefore, the two distributions $f_{g/B}$ and $h_{g/B}$
depend on the parameter $\zeta_v$ which is defined by replacing in
$\zeta_u$ $P_A$ with $P_B$ and $u$ with $v$ in Eq. (\ref{Zeta}).

\par
To study the TMD factorization of the process in Eq. (\ref{proc}),
we need to study
\begin{equation}
 g(p) + g(\bar p) \to \eta_Q(q) +X ,
\label{gg}
\end{equation}
with $p^\mu =(P^+_A,0,0,0)$ and $\bar p^\mu =(0,P_B^-,0,0)$. Since
we are interested in the kinematical region of the small transverse
momentum, we need to study the process in the limit of $q_\perp \ll
Q=M_{\eta_Q}$. In reality, initial hadrons are bound states of
partons. One can imagines that $\eta_Q$ can be produced through
two-gluon fusion, as in Eq. (\ref{gg}), in which one gluon is from
the hadron $h_A$ and another is from the hadron $h_B$. Certainly,
there can be interactions or gluon exchanges between spectators in
$h_A$ and those in $h_B$ and between partons involved in Eq.
(\ref{gg}) and spectators. If these interactions are of short
distances or if the exchanged gluons are hard, their effects in
cross sections can be factorized with operators of higher twists
because that the involved processes are a scattering of
multipartons. These effects are power suppressed and can be
neglected. A factorization may not be obtained if the interactions
are of long distance or if the exchanged gluons are soft. It has
been shown in Drell-Yan processes \cite{CSS,JCSOST} that the effects
of soft-gluon exchanges are canceled or power suppressed if the sum
of the unobserved states is completed. The exchanged gluons can be
those collinear to the initial hadron $h_A$ or $h_B$; the effects of
these collinear gluons can be factorized into the gauge links in the
corresponding parton distribution functions. Since the process in
Eq. (1) is similar to Drell-Yan processes, we expect that the
conclusion made in \cite{CSS,JCSOST} for Drell-Yan processes also
applies here. In our case, we have an observed $\eta_Q$ in the final
state. In general, $\eta_Q$ is a bound state of a heavy-quark pair
and possible light partons. We will use NRQCD for $\eta_Q$. In the
approximation explained later, $\eta_Q$ is effectively taken as a
$Q\bar Q$ state in which the state is in color singlet and there is
no relative momentum between $Q$ and $\bar Q$. This $Q\bar Q$ is
effectively pointlike and cannot emit soft gluons. Hence, there are
no soft interactions between $\eta_Q$ and spectators at leading
power. With the arguments given in the above, we only need to study
the process in Eq. (\ref{gg}) for factorization.
\par
The reason why we only need to study the process in Eq. (\ref{gg})
at the leading power for the factorization can be understood in
another way: If the factorization holds or is proven, it holds for
arbitrary hadrons in the initial state. Especially, it also holds if
the initial states are of partons. In the case with a factorization
which is not rigorously proven, one can use parton states to study
or to examine it, and to eventually prove it. In this work, we use
the process in Eq. (\ref{gg}) to study the relevant factorization
beyond tree level.
\par
For long-distance effects related to $\eta_Q$, we use NRQCD
factorization. We will work at the leading order of the small
velocity expansion in NRQCD. At this order, the production of
$\eta_Q$ can be thought as a two-step process. In the first step, a
$Q\bar Q$ pair is produced in which the heavy quark $Q$ and its
antiquark $\bar Q$ carry the same momentum $q/2$. The pair is in
color-singlet and spin-singlet $^1 S_0$. Then, the pair is
transmitted into $\eta_Q$ with the mass $Q=2m_Q=M_{\eta_Q}$. The
transition is described by a NRQCD matrix element. It is noted that
the considered $Q\bar Q$ pair is in color singlet and hence there is
no interaction of long distance between the $Q\bar Q$ pair and
spectators of initial hadrons, as discussed before. At higher orders
of the small velocity expansion the $Q\bar Q$ pair can be in the
color-octet state \cite{nrqcd}. With the color-octet $Q\bar Q$ pair
it is possible that the NRQCD factorization proposed in \cite{nrqcd}
is violated beyond the one-loop level indicated by the study in
\cite{NQS}.

\par
At tree level, the process in Eq. (\ref{gg}) is with $X$ as nothing.
It is straightforward to obtain the differential cross section,
\begin{eqnarray}
 \frac {d \sigma } { d x d y d^2 q_\perp}
   &=&  \sigma_0 \frac{\pi}{ Q^2 }\delta (xy s -Q^2)
  \delta (1-x) \delta (1-y) \delta^2 (\vec q_\perp) ,
\nonumber\\
   &&   \sigma_0 =\frac{(4\pi \alpha_s)^2 }{N_c (N_c^2-1) m_Q} \vert \psi (0) \vert^2,
\label{tree}
\end{eqnarray}
with $s = 2 p^+ \bar p^-$ and $m_Q$ being the pole mass of the heavy
quark. $\psi(0)$ is the wave function of $\eta_Q$ at the origin. In
fact, $ \vert \psi (0) \vert^2$ should be expressed as a NRQCD
matrix element. Beyond tree level, Coulomb singularities
representing long-distance effects related to $\eta_Q$ appear. These
singularities are factorized into the NRQCD matrix element. At tree
level, one easily finds
\begin{equation}
  f^{(0)}_{g/A} (x,k_\perp, \zeta^2_u,\mu) =f^{(0)}_{g/B} (x,k_\perp, \zeta^2_v,\mu) =\delta(1-x)
  \delta^2 (\vec k_\perp),
\label{TMDg}
\end{equation}
while $h_{g/A}$ and $h_{g/B}$ are zero. They become nonzero at order
of $\alpha_s$. With these results, one can write the tree-level
cross section as a factorized form
\begin{eqnarray}
\frac {d \sigma} { d x d y d^2 q_\perp} &=& \frac{\pi\sigma_0}{ Q^2 } \int d^2k_{a\perp} d^2 k_{b\perp}
 f_{g/A} (x,k_{a\perp}) f_{g/B} (y,k_{b\perp}) \delta^2 (\vec k_{a\perp}
  + \vec k_{b\perp} -\vec q_\perp) \delta (xy s - Q^2) {\mathcal H},
\nonumber\\
 {\mathcal H} &=& 1 +{\mathcal O}(\alpha_s).
\label{Tree-Fac}
\end{eqnarray}
Beyond the tree level, one needs to introduce a soft factor. As we
will see explicitly , all soft divergences will be factorized into
the soft factor and TMD gluon distributions so that the perturbative
coefficient ${\mathcal H}$ is free from soft divergence. We will
then determine ${\mathcal H}$ at one-loop level.
\par
\par
\begin{figure}[hbt]
\begin{center}
\includegraphics[width=10cm]{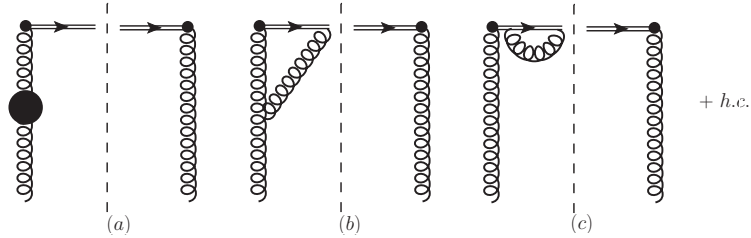}
\end{center}
\caption{The one-loop corrections to the gluon TMD. The double lines
represent the gauge link. The black bubble in Fig. 1a is for
self-energy correction.  } \label{P2}
\end{figure}
\par
To derive the factorization at one loop, we need to study the
one-loop corrections to TMD gluon distributions and the differential
cross section. The one-loop correction to TMD gluon distribution has
been studied in \cite{JMYG}, where the collinear divergence has been
regularized with an infinitely small off-shellness of the gluon.
Here, we regularize all divergences in $d=4-\epsilon$ space-time.
The correction can be divided into the virtual and real corrections.
The virtual correction is given by diagrams in Fig. 1. We will use
the $\overline{\rm MS}$ scheme to subtract ultraviolet (UV)
divergences. After the subtraction, we have the virtual correction
from Fig. 1,
\begin{eqnarray}
f_{g/A}^{(1)}(x,k_\perp,\zeta_u,\mu )\biggr\vert_{vir.} &=& \frac{\alpha_s}{4\pi} \delta (1-x) \delta^2 (\vec k_\perp)
\left [
 \left ( -\frac{2}{\epsilon_{s}} + \ln \frac{e^\gamma \mu^2}{4\pi \mu_s^2} \right )
       \left ( \frac{11}{3} N_c -\frac{2}{3} N_F \right )   \right.
\nonumber\\
  && \left.  + 2 N_c \left  ( -\frac{4}{\epsilon_s^2}
   -\frac{2}{\epsilon_s} \ln \frac{4\pi\mu_s^2}{e^\gamma \zeta_u^2}
    - \frac{1}{2} \ln^2 \frac{4\pi\mu_s^2}{e^\gamma \zeta_u^2} -\frac{5\pi^2}{12}
   + \left ( -\frac{2}{\epsilon_s} +\ln\frac{e^\gamma\zeta_u^2}{4 \pi \mu_s^2} \right )
\right. \right.
\nonumber\\
  && \left. \left.  + \frac{1}{2} \ln \frac{\mu^2}{\zeta_u^2} -\frac{3}{2} \right ) \right ] ,
\end{eqnarray}
where the poles in $\epsilon_s =4-d$ stand for collinear or infrared
divergences, i.e., soft divergences. $\mu_s$ is the scale associated
with these poles. $\mu$ is the UV scale. The terms in the first line
in Eq. (10) is the sum of the contributions from Fig. 1a and Fig. 1c
with their conjugated diagrams. The remaining terms are from Fig. 1b
and its conjugated diagram.

\par
\begin{figure}[hbt]
\begin{center}
\includegraphics[width=11cm]{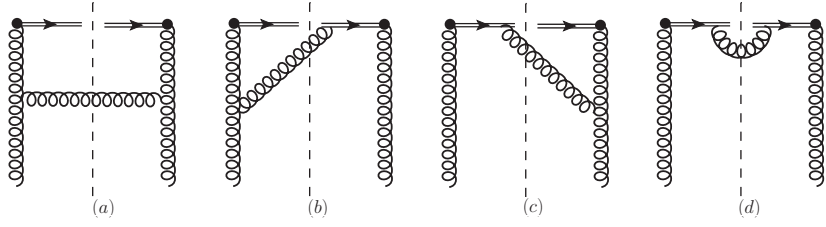}
\end{center}
\caption{The real correction at one loop to the gluon TMD. The
double lines represent the gauge link. These diagrams are for real
corrections.   } \label{P2}
\end{figure}
\par

The corrections from Fig. 2 are real corrections. They can be found
in \cite{JMYG} as
\begin{eqnarray}
f_{g/A}^{(1)}(x,k_\perp,\zeta_u,\mu )\biggr\vert_{re.} &=& \frac{\alpha_s N_c}{\pi^2 k^2_\perp} \left [ \left ( \frac{1-x}{x}
  + x(1-x) +\frac{x}{2} \right ) -\frac{1}{2}\delta(1-x)
\right.
\nonumber\\
  && \left. \ \ \ \ \ \ \ \ \  +   \frac{x}{(1-x)_+}
  -\frac{x}{2}  + \frac{1}{2} \delta (1-x) \ln\frac{\zeta_u^2}{k_\perp^2} \right ] ,
\end{eqnarray}
where the terms in the first line are from Fig. 2a and Fig. 2d. The
total one-loop correction is then the sum of the virtual and real
corrections. At one loop, $h_{g/A}$ becomes nonzero. It receives a
contribution from Fig. 2a. We have
\begin{equation}
   h_{g/A}(x,k_\perp,\zeta_u,\mu) = \frac{2 \alpha_s N_c}{\pi^2 (k_\perp^2)^2}\frac{1-x}{x} + {\mathcal O}(\alpha_s^2).
\end{equation}
By replacing $\zeta_u$ with $\zeta_v$ we obtain $f_{g/B}$ and $h_{g/B}$ from $f_{g/A}$ and $h_{g/A}$, respectively.

\par
\begin{figure}[hbt]
\begin{center}
\includegraphics[width=9cm]{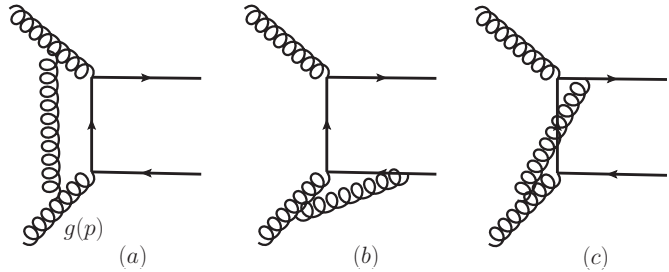}
\end{center}
\caption{The class of diagrams where a gluon is emitted from the
initial gluon $g(p)$ and is attached to a possible place. There are
six diagrams. Three of them are given here. Another three diagrams
are obtained by reversing the direction of the heavy quark line. }
\label{P1}
\end{figure}
\par
Now we turn to one-loop corrections of the differential cross
section. The corrections can be divided into the virtual correction
and the real correction. The virtual correction is the one-loop
correction to the process $g(p) + g(\bar p)\to \eta_Q(q)$. We denote
the total contribution from the virtual correction as
\begin{eqnarray}
 \frac {d \sigma (gg\to \eta_Q) } { d x d y d^2 q_\perp}\biggr\vert_{vir.} =  \frac{1}{2 s (2\pi)^3 } \delta ( xy s -Q^2)
 \delta (1-x)\delta (1-y) \delta^2(\vec q_\perp)  \sigma_1.
\end{eqnarray}
The contributions to $\sigma_1$ can be divided into four parts,
\begin{equation}
\sigma_1 = \sigma_{1A} + \sigma_{1B} + \sigma_{1C} +\sigma_{1D},
\end{equation}
$\sigma_{1A}$ receives contributions from diagrams in which a
virtual gluon is emitted by the initial gluon $g(p)$. The diagrams
for this part are given in Fig. 3. $\sigma_{1B}$ receives
contributions from diagrams in which a virtual gluon is emitted by
the initial gluon $g(\bar p)$. $\sigma_{1C}$ denotes the
contributions from diagrams in which a virtual gluon is exchanged
between heavy quark line. $\sigma_{1D}$ denotes the one-loop
corrections of external gluon lines. This part will not contribute
to ${\mathcal H}$, because the contributions to $\sigma_{1D}$ are
automatically subtracted into TMD gluon distributions. Below, we
will only give and discuss the results of $\sigma_{1A,1B,1C}$.
\par
In the above classification, Fig. 3a can contribute both to
$\sigma_{1A}$ and $\sigma_{1B}$. We put the half of the contribution
Fig. 3a into $\sigma_{1A}$ and another half into $\sigma_{1B}$. With
symmetry arguments one easily finds $\sigma_{1A}=\sigma_{1B}$. We
have then
\begin{eqnarray}
\sigma_{1A} =  \sigma_{1B} = \frac{1}{2}\sigma_1\biggr\vert_{3a} +\sigma_1\biggr\vert_{3b} +\sigma_1\biggr\vert_{3c}.
\end{eqnarray}
By summing contributions from each diagram we
obtain the following results for the virtual corrections:
\begin{eqnarray}
\frac{\sigma_{1A}}{\sigma_0} &=& \frac{\alpha_sN_c }{12 \pi} \biggr [ -6 \frac{4}{\epsilon_s^2} -6 \frac{2}{\epsilon_s}
\biggr ( 1 + \ln \frac{e^{-\gamma} 4\pi \mu_s^2}{Q^2} \biggr )  -3\ln ^2\frac{e^{-\gamma} 4\pi \mu_s^2}{Q^2}
-6\ln \frac{e^{-\gamma} 4\pi \mu_s^2}{Q^2}
\nonumber\\
  &&  + 9 \ln\frac{\mu^2}{Q^2} -6\ln 2 + 6 +\frac{11}{4} \pi^2 \biggr ],
\nonumber\\
\frac{\sigma_{1C}}{\sigma_0}
    &=& \frac{\alpha_s}{2 \pi} \biggr [  -N_c \ln \frac{\mu^2}{Q^2} + C_F   \biggr (
   -2 + 4 \ln 2 \biggr )
+\frac{1}{N_c} \biggr (
    2 \ln 2 -\frac{1}{4}\pi^2 \biggr )  \biggr ].
\end{eqnarray}
In these results, the UV poles are subtracted in the ${\overline
{\rm MS}}$ scheme. The on-shell scheme for the renormalization of
heavy quark propagators is used so that $m_Q=Q/2$ is the pole mass
of heavy quark. In $\sigma_{1A}$, the pole terms of $\epsilon_s$ are
for soft divergences coming only from Fig. 3a. The contributions
from Fig. 3b and Fig. 3c also contain collinear divergences and
infrared divergences. The infrared divergences are canceled in the
sum of the two diagrams, because the $Q\bar Q$ is in color singlet.
The collinear divergences are also canceled. In calculating the
diagrams for $\sigma_{1C}$ one will meet Coulomb singularity. This
singularity is factorized into NRQCD matrix element. Hence, we have
finite $\sigma_{1C}$.
\par
The real correction is from the tree-level process
\begin{equation}
   g(p) + g(\bar p) \to \eta_Q (q) + g(k).
\label{Real}
\end{equation}
For the color-single $Q\bar Q$ pair, there are 12 diagrams for the
amplitude. Since we are interested in the low $q_\perp$ region, we
expand the differential cross section in $q_\perp/ Q$ and only take
the leading order in the expansion. At the leading order, we have
only those diagrams given in Fig. 4 for the differential cross
section. The result for the process in Eq. (\ref{Real}) in the limit
of $q_\perp\to 0$ is
\begin{eqnarray}
\frac{d\sigma}{dx dy d^2q_\perp} &=& \frac{\pi \sigma_0}{Q^2}
    \frac{N_c \alpha_s}{4 \pi^2  q^2_\perp} \delta (xy s- Q^2)  \left [  \frac{2 \delta (1-y) }{ x }
       \biggr ( 2 -2 x + 3 x^2 -2 x^3 \biggr )
 +  \frac{x (1+x)}{(1-x)_+} \delta (1-y)
\right.
\nonumber\\
       && \left. \ \  -  \delta (1-x)\delta (1-y) \ln \frac{q_\perp^2}{Q^2} + ( x\leftrightarrow y )  \right ]   + {\mathcal O}(q_\perp^0).
\end{eqnarray}

\par
\begin{figure}[hbt]
\begin{center}
\includegraphics[width=13cm]{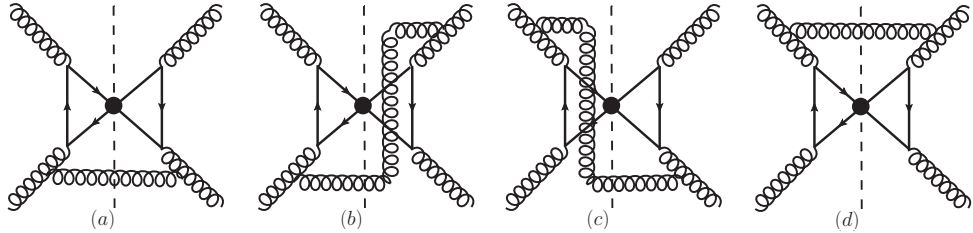}
\end{center}
\caption{The diagrams for the cross section of $g+ g\to g +\eta_Q$.
In these diagrams, the gluon in the intermediate state is emitted or
absorbed by gluons. The black dots denote the projection of the
$Q\bar Q$ pair into the color singlet $^1S_0$ state. By reversing
the quark lines, one can obtain other three diagrams from each
diagram. } \label{P2}
\end{figure}
\par

\par
The factorized result in Eq. (\ref{Tree-Fac}) is derived at tree
level. If we extend the factorization beyond tree level, with the
one-loop results, in the above we will find the following: (i) The
soft divergences are not factorized, i.e., ${\mathcal H}$ will
contain some infrared divergences represented by poles in
$\epsilon_s$.  (ii) The real correction of the differential cross
section is not totally generated by TMD gluon distributions. In
other words, ${\mathcal H}$ will receive correction from the real
correction. It results in that ${\mathcal H}$ depends on $q_\perp$.
All of these have a common reason. In the one-loop corrections to
the differential cross section, there is an exchange of a soft gluon
between the two initial gluons $g(p)$ and $g(\bar p)$.  In the
virtual correction, the exchange results in infrared divergences,
and in the real correction it results in contributions proportional
to $\delta(1-x)\delta(1-y)$. The effects of the soft gluon exchange
are not exactly generated by the corresponding soft gluon exchange
in TMD gluon distributions. The effects of soft gluon exchange are
of long distance. Therefore, one needs to introduce a soft factor in
the factorization to completely factorize these effects from
${\mathcal H}$ determined with Eq. (\ref{Tree-Fac}).
\par
The effects of soft gluon exchange between a gluon moving in the $+$
direction and a gluon moving in the $-$ direction can be described
by the expectation value of a product with four gauge links. We
introduce, as in \cite{JMYG},
\begin{eqnarray}
S(\vec b_\perp,\mu,\rho) = \frac{1}{N_c^2-1} \langle 0\vert {\rm Tr} \left  [   {\mathcal L}^\dagger_v (\vec b_\perp,-\infty)
  {\mathcal L}_u (\vec b_\perp,-\infty) {\mathcal L}_u^\dagger (\vec 0,-\infty){\mathcal L}_v (\vec 0,-\infty)   \right ] \vert 0\rangle.
\label{SoftS}
\end{eqnarray}
The gauge links are past pointing. It reflects the fact that the two
gluons $g(p)$ and $g(\bar p)$ are in the initial state. The
dependence on the directions of gauge links is only through the
parameter $\rho^2=(2 u\cdot v)^2/(u^2 v^2) \approx u^-v^+/(u^+v^-)$.
The limits $u^-\gg u^+$ and $v^+\gg v^-$ are taken similarly to that
in TMD gluon distributions. The gauge links or the gauge field is in
the adjoint representation. At leading order, one has
\begin{equation}
S^{(0)}(\vec b_\perp,\mu,\rho) =1.
\end{equation}
\par
\begin{figure}[hbt]
\begin{center}
\includegraphics[width=11cm]{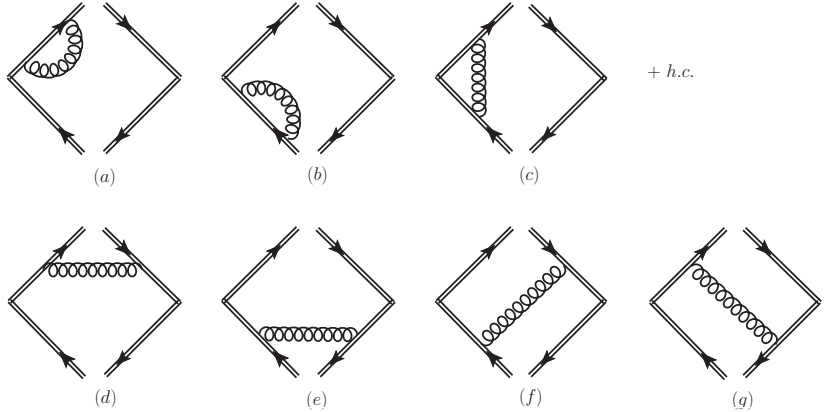}
\end{center}
\caption{One-loop corrections for the soft factor. The first three diagrams plus their
complex conjugated are virtual corrections. The last four diagrams are real corrections.
A cut line is implied.}
\label{soft}
\end{figure}
\par
At one loop, there are corrections from Fig. 5. One can divide the
corrections into a virtual and a real part. The diagrams in the
first row are of the virtual part. Those in the second row are of
the real part. The virtual correction reads
\begin{eqnarray}
S^{(1)}_{vir.} (\vec b_\perp,\mu,\rho) = \frac{\alpha_s N_c}{2\pi} \left [ -\frac{2}{\epsilon_s}
  + \ln\frac{e^\gamma \mu^2}{4 \pi \mu_s^2} \right ] \left ( 2 -\ln\rho^2 \right ),
\end{eqnarray}
where the UV pole is subtracted. The pole in $\epsilon_s$ represents
the IR divergence with the scale $\mu_s$. The real part is
\begin{eqnarray}
S^{(1)}_{re.} (\vec b_\perp,\mu,\rho) = -\frac{\alpha_s N_c}{2\pi^2}
 \left ( 2 -\ln\rho^2 \right ) \int d^2 k_\perp
 \frac{e^{-i\vec b_\perp\cdot \vec k_\perp}}{ k_\perp^2}.
\end{eqnarray}
The total one-loop contribution is the sum of the virtual and real
contributions.
\par
We now define our soft factor which will enter the TMD factorization
as
\begin{eqnarray}
\tilde S(\vec\ell_\perp,\mu,\rho) &&= \int\frac{d^2 b_\perp}{(2\pi)^2} e^{ i\vec b_\perp\cdot \vec\ell_\perp}
  S^{-1}(\vec b_\perp,\mu,\rho)
\nonumber\\
    && = \delta^2(\vec\ell_\perp) -\frac{\alpha_s N_c}{2\pi} \left ( 2 -\ln\rho^2 \right )
     \left [ \left ( -\frac{2}{\epsilon_s} +\ln\frac{ e^\gamma \mu^2}{4\pi \mu_s^2} \right )
       \delta^2 (\vec\ell_\perp) - \frac{1}{\pi \ell^2_\perp} \right ] +{\mathcal O}(\alpha_s^2).
\label{SoftS1}
\end{eqnarray}
With the introduced soft factor, we propose the TMD factorization as
\begin{eqnarray}
\frac {d \sigma} { d x d y d^2 q_\perp} &=& \frac{\pi \sigma_0}{Q^2 } \int d^2k_{a\perp} d^2 k_{b\perp}
d^2\vec \ell_\perp  \delta^2 (\vec k_{a\perp} + \vec k_{b\perp}+\vec\ell_\perp -\vec q_\perp)  \delta (xy s -Q^2)
\nonumber \\
   && \ \ \ \ \ \  \cdot  f_{g/A} (x,k_{a\perp},\zeta_u, \mu) f_{g/B} (y,k_{b\perp},\zeta_v,\mu) \tilde S(\ell_\perp, \mu,\rho)   {\mathcal H}(Q,\mu, \zeta_u,\zeta_v).
\label{S-Fac}
\end{eqnarray}
From one-loop results of the differential cross section, TMD gluon
distributions, and the soft factor, we derive
\begin{eqnarray}
{\mathcal H}(Q,\mu, \zeta_u,\zeta_v) &=& 1 +
\frac{\alpha_s N_c}{4\pi}
 \left [  \ln^2 \frac{\zeta_u^2}{Q^2}+\ln^2\frac{\zeta_v^2}{Q^2} -\ln\rho^2\left (1+ 2 \ln\frac{\mu^2}{Q^2} \right )
  + 2 \ln\frac{\mu^2}{Q^2}  + \frac{7}{2} \pi^2
\right.
\nonumber\\
  && \left.  +\frac{2}{N_c^2} \biggr (
    1 -\frac{1}{4}\pi^2 \biggr )  \right ]
   + {\mathcal O}(\alpha_s^2).
\label{Fac}
\end{eqnarray}
It is clear that ${\mathcal H}$ is free from any soft divergence and
does not depend on $q_\perp$. With the factorization the small
transverse momentum, $q_\perp$ is generated by the transverse motion
of gluons in the initial hadrons and by soft gluon radiation.
Equations (\ref{S-Fac}) and (\ref{Fac}) are our main results. It
should be noted that the factorization holds for arbitrary large
$\zeta_u$ and $\zeta_v$. For practical applications, one may take a
frame to simplify the results in Eqs. (\ref{S-Fac}) and (\ref{Fac}).
One can take $\zeta_u^2 =\zeta_v^2 = \rho Q^2$ so that the TMD gluon
distributions in Eq. (\ref{S-Fac}) depend on $\rho$ and $Q^2$ and
the perturbative coefficient becomes a function of $Q$, $\mu$, and
$\rho$,
\begin{eqnarray}
{\mathcal H}(Q,\mu, \rho) = 1 +
\frac{\alpha_s N_c}{2\pi}
 \left [  \ln^2 \rho  - \ln\rho\left (1+ 2 \ln\frac{\mu^2}{Q^2} \right )
  +  \ln\frac{\mu^2}{Q^2}   + \frac{7}{4} \pi^2
  +\frac{1}{N_c^2} \biggr (
    1 -\frac{1}{4}\pi^2 \biggr )  \right ]
   + {\mathcal O}(\alpha_s^2).
\label{Fac1}
\end{eqnarray}

\par
\begin{figure}[hbt]
\begin{center}
\includegraphics[width=7cm]{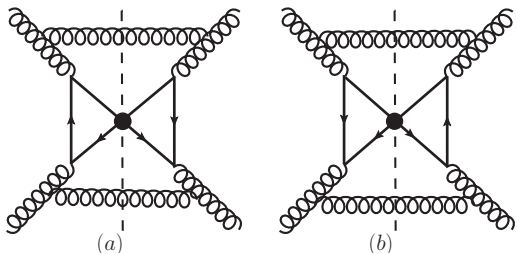}
\end{center}
\caption{The diagrams for the cross section of $g+ g\to g+g+\eta_Q$
give the contributions factorized with the gluon TMD $g_k$. By
reversing the quark lines, one can obtain other three diagrams from
each diagram.     } \label{P2}
\end{figure}
\par
At the considered orders, we will not find the contributions which
can be factorized with $h_{g/A}$ or $h_{g/B}$ defined in Eq.
(\ref{DEF}). However, there is a contribution involving these
distributions of linearly polarized gluons in the TMD factorization.
This contribution can be found at a higher order of $\alpha_s$ from
diagrams given in Fig. 6. It is straightforward to calculate these
diagrams in the limit $q_\perp \to 0$. We find that the contribution
takes the factorized form
\begin{eqnarray}
\frac {d \sigma} { d x d y d^2 q_\perp}\biggr\vert_{Fig.6}  &=&  \frac{\pi\sigma_0}{ Q^2 } \delta (xy s - Q^2) \int d^2k_{a\perp} d^2 k_{b\perp}
  \delta^2 (\vec k_{a\perp}
  + \vec k_{b\perp} -\vec q_\perp)  \biggr [ f_{g/A} (x,k_{a\perp}) \biggr\vert_{2a}
   f_{g/B} (y,k_{b\perp}) \biggr\vert_{2a}
\nonumber\\
   && \ \ \ \  - \frac{1}{2} \biggr ( (\vec k_{a\perp}\cdot \vec k_{b\perp})^2 -\frac{1}{2} k^2_{a\perp} k^2_{b\perp} \biggr )  h_{g/A} (x,k_{a\perp}) h_{g/B}  (y,k_{b\perp})  \biggr ].
\end{eqnarray}
Our result in the last line has also  been derived in \cite{BoPi}
with a different method. The perturbative coefficient of the
contribution in the last line is at order of $\alpha_s^0$. This
contribution should be added to Eq. (\ref{S-Fac}). In principle one
can determine the perturbative coefficient of the contribution
beyond the leading order of $\alpha_s$ following the same way as has
been done for Eqs. (\ref{S-Fac}) and (\ref{Fac}). However, this will
be very tedious because one needs to calculated the partonic process
$g+g\to \eta_Q +X$ at the 3-loop level. We leave this for a future
study.
\par
In the factorized form of the differential cross section in Eq.
(\ref{Fac}), the TMD gluon distributions do not depend on processes,
they only depend on hadrons. The perturbative coefficient ${\mathcal
H}$ does not depend on initial hadrons. The soft factor $\tilde S$
defined in Eqs. (\ref{SoftS}) and (\ref{SoftS1}) is a basic quantity
of QCD, i.e., it depends neither on hadrons or on processes. It is
noted that the same soft factor also appears in TMD factorization of
Higgs production studied in \cite{JMYG}. This indicates that soft
divergences in different processes or in a class of processes can be
factorized into the same object. This implies that the soft factor
is universal at certain level. In TMD factorization of Drell-Yan
processes, one also needs a soft factor to take radiation of soft
gluons to complete the factorization\cite{CSS,JMY1}. The soft factor
there is similar to that defined in Eqs. (\ref{SoftS}) and
(\ref{SoftS1}). The only difference is that they are defined in
different $SU(3)$ representations.
\par
The studied TMD factorization can be used for the region with
$q_\perp\sim \Lambda_{QCD}$ for extracting TMD gluon distributions.
However, its usage is not limited to this kinematic region, because
the factorization holds, in general, in the region $q_\perp/Q \ll
1$. In the region $Q\gg q_\perp \gg \Lambda_{QCD}$, both TMD
factorization and collinear factorization hold. In the collinear
factorization, the perturbative coefficient functions in this region
contain large log of $q_\perp/Q$. The results from TMD factorization
can be used to resum these large logs. This leads to the well-known
Collins-Soper-Sterman resummation \cite{CSS}. Based on our result
here, one can also derive the resummation in the case of quarkonium
production, similarly to that derived in \cite{JMYG}. We, therefore,
do not discuss the details about the resummation here. We note that
such a resummation has been studied  very recently in \cite{SYY}. An
early work about the resummation can be found in \cite{BQW}, where
the formation of a quarkonium from a $Q\bar Q$ pair is described
with a color evaporation model instead of NRQCD factorization.
\par
At the orders we have considered, the production of a p-wave
quarkonium is possible. However, the TMD factorization in this case
can be complicated. According to the NRQCD factorization in
\cite{nrqcd}, one needs to consider not only the contribution from
the production of a color singlet p-wave $Q\bar Q$ pair, but also
the contribution of a color-octet s-wave $Q\bar Q$ pair. The
formation of a p-wave quarkonium from the color-singlet and the
color-octet $Q\bar Q$ pair is at the same order in the small
velocity expansion. In the case we studied here, we only need to
consider the contribution from production of a color-singlet s-wave
$Q\bar Q$ pair. At the leading power the pair decouples with soft
gluons. However, in the case of p-wave quarkonia, the color-singlet
p-wave and color-octet s-wave $Q\bar Q$ pair can emit soft gluons at
leading power. To completely separate the effects of soft gluons,
one may need a different soft factor than that introduced here. This
is also the reason why we write our TMD factorization in Eq.
(\ref{S-Fac}) explicitly with the unsubtracted TMD gluon
distributions and the soft factor. Another complication with p-wave
quarkonia is that one needs a gauge link for the NRQCD matrix
element of the contribution from the color-octet $Q\bar Q$ pair to
establish NRQCD factorization beyond one loop, as shown in
\cite{NQS}. We will examine the TMD factorization for p-wave
quarkonium in a separate publication.
\par
Before summarizing our work, we note that one can define subtracted
TMD gluon distributions  as those used for Higgs production in
\cite{JMYG}, to factorize the differential cross section. Then, our
result can be factorized as the same form in \cite{JMYG} only with
the difference that the perturbative coefficient is different. One
may also redefine TMD gluon distributions as suggested in \cite{JC1}
so that the differential cross section is factorized only with the
redefined TMD gluon distributions. Due to this and the reason
discussed for p-wave quarkonium, we only give our results factorized
with the unsubtracted TMD gluon distributions as in Eq.
(\ref{S-Fac}).

\par
To summarize, we have studied the one-loop TMD factorization of
$^1S_0$-quarkonium production in a hadron collision at low
transverse momentum. We find that the differential cross section can
be factorized with the TMD gluon distributions, the soft factor, and
the perturbative coefficient. The TMD gluon distributions and the
soft factor are consistently defined with QCD operators; the
perturbative coefficient is determined here at one loop. In
comparison with the factorization derived at tree level, the soft
factor is needed to cancel all effects of soft gluons. Our result
will be useful not only for extracting TMD gluon distributions from
experimental data, but also for resumming large logs of $q_\perp$
appearing in the collinear factorization.

\par\vskip20pt
\par

\par\vskip40pt
\noindent
{\bf Acknowledgments}
\par
We would like to thank Prof. Y.-N. Gao for a discussion about LHCb
experiment and G. P. Zhang for discussions about TMD parton
distributions. The work of J. P. M. is supported by National Nature
Science Foundation of People's Republic of China (Grants No.
10975169, 11021092, and No. 11275244). The work of J. X. W. is
supported by the National Natural Science Foundation of Peoples
Republic of China (Grants No. 10979056 and No. 10935012), and in
part by DFG and NSFC (CRC 110).
\par\vskip40pt

\par

\end{document}